\documentstyle[aps,psfig]{revtex}
\begin{document}
\draft
\title{In-medium vector meson properties and low-mass dilepton
production from hot hadronic matter}

\author{Amruta Mishra\footnote[1]
{email: mishra@th.physik.uni-frankfurt.de}, 
Joachim Reinhardt, Horst St\"ocker 
and Walter Greiner}
\address {Institut f\"ur Theoretische Physik, 
J.W. Goethe Universit\"at, Robert Mayer-Stra{\ss}e 10,\\
Postfach 11 19 32, D-60054 Frankfurt/Main, Germany}

\maketitle
\begin{abstract}
The in-medium properties of the vector mesons are known to be modified
significantly in hot and dense hadronic matter due to vacuum
polarisation effects from the baryon sector in the Walecka model.  The
vector meson mass drops significantly in the medium due to the effects
of the Dirac sea. In the variational approach adopted in the present
paper, these effects are taken into account through a realignment of
the ground state with baryon condensates. Such a realignment of the
ground state becomes equivalent to summing of the baryonic tadpole
diagrams in the relativistic Hartree approximation (RHA).  The
approximation scheme adopted here goes beyond RHA to include quantum
effects from the scalar meson and is nonperturbative and
self--consistent. It includes multiloop effects, thus corresponding to
a different approximation as compared to the one loop approximation of
including scalar field quantum corrections.  In the present work, we
study the properties of the vector mesons in the hot and dense matter
as modified due to such quantum correction effects from the baryon as
well as scalar meson sectors. These medium modifications of the
properties of the vector mesons are reflected, through the shifting
and broadening of the respective peaks, in the low mass dilepton spectra. 
There is broadening of the peaks due to corrections from scalar meson
quantum effects as compared to the relativistic Hartree
approximation. It is seen to be rather prominent for the $\omega$
meson in the invariant mass plot.
\end{abstract}

\def\bfm#1{\mbox{\boldmath $#1$}}

\section{Introduction}
The in-medium properties of the vector mesons ($\rho$ and $\omega$)
in hot and dense matter have an important role to play in 
the low mass dilepton production resulting from heavy ion collision 
experiments. This has, hence, been a topic of great interest 
in the recent past, both experimentally \cite {helios,ceres,dls,rhic,hades} 
and theoretically \cite{brown,hat,jin,samir,weise,ernst}. 
The experimental observation of enhanced dilepton production
\cite{helios,ceres,dls} in the low 
invariant mass regime could be due to a reduction in the vector 
meson masses in the medium. Brown and Rho suggested 
the hypothesis that the vector meson masses drop in the medium
according to a simple (BR) scaling law \cite{brown}, given as 
$m_V^*/m_V=f_\pi^*/f_\pi$, where $f_\pi$ is the pion decay constant 
and asterisk refers to in-medium quantities. 
In the  framework of Quantum Hadrodynamics (QHD) as a description of
the hot hadronic matter, it is seen that
the dropping of the vector meson masses has its dominant
contribution arising from the vacuum polarisation effects from the
baryon sector \cite{hatsuda,hatsuda1,jeans,sourav},
which is not observed in the mean field approximation.
This means that the quantum effects do play an important role in
the medium modification of vector meson properties.
There have also been approaches based on QCD sum rules \cite {hat,jin}
which confirm a scaling hypothesis \cite {hat} as suggested by Brown 
and Rho, with a saturation
scheme that leads to a delta function at the vector meson pole and
a continuum for higher energies for the hadronic spectral function.
It is, however, seen that such a simple saturation scheme for finite
densities does not work and a more realistic description for the
hadronic spectral function is called for \cite{samir}. Using an
effective Lagrangian model to calculate the hadronic spectral 
function, it is seen that such a universal scaling law is not 
suggested for in-medium vector meson masses \cite{weise}.

The medium modifications of the vector mesons have been thus a subject
of several investigations. It has been emphasized in the literature
\cite {sourav,wambach} that the properties of the hadrons are modified
due to their interactions with the thermal bath, and such
modifications are reflected in the dilepton and photon spectra emitted
from a hot and dense matter \cite{koch}.  Dileptons are interesting
probes for the study of evolving matter arising from relativistic
heavy ion collisions since they do not interact strongly and escape
unthermalized from the hot and dense matter at all stages of the
evolution. The temperature \cite{temp} and density \cite{dens}
modifications of the dileptons from hot hadronic matter as well as
from a Quark Gluon Plasma (QGP) resulting from heavy ion collisions
have been a subject of extensive investigations in the recent
past. Broadly, two types of modifications for the hadrons in the
medium are expected: shift in the pole position for the hadron
propagator, giving rise to a modification of the mass, and broadening
of the spectral function.  In the present work, we shall attempt to
study the medium modification of vector meson masses and decay widths
in the hot hadronic matter within the framework of Quantum
Hadrodynamics (QHD) taking the vacuum polarisation effects into
account, and their subsequent effect on the dilepton spectra.

In the conventional hadronic models \cite {serot,chin}, the masses of
the vector mesons stay constant or increase slightly, in the mean
field approximation, i.e., when the polarization from the Dirac sea is
neglected \cite{chin}. With the inclusion of quantum corrections from
the baryonic sector, however, in the Walecka model one observes a drop
in the vector meson masses in the medium
\cite{hatsuda,hatsuda1,jeans}.  This medium modification of the vector
meson masses plays an important role in the enhanced dilepton yield
\cite{wambach} for masses below the $\rho$ resonance in the heavy ion
collision experiments \cite{helios,ceres}.  It has been emphasized
recently that the Dirac sea contribution (taken into account through
summing over baryonic tadpole diagrams in the relativistic Hartree
approximation (RHA)) dominates over the Fermi sea contribution and
$m_\omega^*/m_\omega <1$ is caused by a large dressing of $\bar N N$
cloud in the medium \cite{jeans}.  Further quantum effects arising
from the scalar mesons have been considered \cite{serot} along with
the RHA for the baryon sector in the context of vector meson mass
modifications in strange hadronic matter \cite{pal}.  The present
method is a step in that direction of studying medium modifications of
the vector meson properties including such quantum effects from the
scalar mesons in a self--consistent manner, along with those arising
from the baryon sector.

It was earlier demonstrated in a nonperturbative formalism that a
realignment of the ground state with baryon-antibaryon condensates is
equivalent to the relativistic Hartree approximation
\cite {mishra}. The ground state for the nuclear matter was extended
to include sigma condensates to take into account the quantum
correction effects from the scalar meson sector \cite {mishra}. 
Such a formalism includes multiloop effects and is self consistent
\cite{mishra,amhm,pi}. The methodology 
was then generalized to consider hot nuclear matter \cite{hotnm} as
well as to the situation of hyperon-rich dense matter \cite{shm}
relevant for neutron stars. The effect of vacuum polarisations on the
vector meson masses has also been recently studied
\cite {vecmass}. In the present work, we study the effect of
such quantum corrections on the in-medium vector meson properties.
This apart, for studying $\rho$ meson properties, we take into account
the $\rho$-$\pi$-$\pi$ interactions, and their effects on the dilepton
production in the low invariant mass regime. The pionic interactions are
seen to modify the $\rho$ meson mass only marginally, as already
stated in the literature \cite{sourav,gale}. Due to scalar meson
contributions, there is broadening in the $\omega$ and $\rho$ peaks,
as compared to RHA.  This is seen to be quite pronounced for the
$\omega$ meson, which leads to smearing and ultimate disappearance of
the $\omega$ peak at finite densities \cite{weise1}.

We organize the paper as follows. We first briefly recapitulate the
nonperturbative framework used for studying the quantum correction
effects in hot nuclear matter in section II. The medium modification
of the $\omega$ and $\rho$ meson masses and decay widths are
considered in section III. We discuss the effect of these medium 
modifications on the low mass dilepton spectra in section IV.  
In section V, we summarize the results of the
present work and give an outlook.

\section {Quantum Vacuum in hot nuclear matter}

We briefly recapitulate here the vacuum polarisation effects arising
from the nucleon and scalar meson fields in hot nuclear matter in a
nonperturbative variational framework \cite{hotnm}. The method of
thermofield dynamics (TFD) \cite {tfd} is used here to study the
``ground state" (the state with minimum thermodynamic potential) at
finite temperature and density within the Walecka model with a quartic
scalar self interaction. The temperature and density dependent baryon
and sigma masses are also calculated in a self-consistent manner in
the formalism. The ansatz functions involved in such an approach are
determined through functional minimisation of the thermodynamic
potential.

The Lagrangian density in the Walecka model is given as
\begin{eqnarray}
{\cal L}&=&\bar \psi \left(i\gamma^\mu \partial_\mu
-M-g_\sigma \sigma-g_\omega\gamma^\mu \omega_\mu\right)\psi
+\frac{1}{2}\partial^\mu\sigma
\partial_\mu\sigma-\frac{1}{2} m_\sigma ^2 \sigma^2
-\lambda \sigma^4
\nonumber\\
&& +\frac{1}{2} m_\omega^2 \omega^\mu \omega_\mu
-\frac{1}{4}(\partial_\mu \omega_\nu -\partial_\nu \omega_\mu)
(\partial^\mu \omega^\nu -\partial^\nu \omega^\mu).
\end{eqnarray}
In the above, $\psi$, $\sigma$, and $\omega_\mu$ are the fields for
the nucleon, $\sigma$, and $\omega$ mesons with masses $M$,
$m_\sigma$, and $m_\omega$ respectively.  Since we are interested in
symmetric nuclear matter, the isovector rho meson does not contribute.
The quartic coupling term in $\sigma$ is necessary for the sigma
condensates to exist, through a vacuum realignment \cite{mishra}.  Our
calculations thus include the quantum effects arising from the sigma
meson in addition to the mean field contribution from the the quartic
self interaction of the scalar meson.  We retain the quantum nature of
both the nucleon and the scalar meson fields, whereas the vector
$\omega$ meson is treated as a classical field, using the mean field
approximation for the $\omega$ meson, given as $\langle \omega^\mu
\rangle=\delta_{\mu 0} \omega_0$. The reason is that without
higher--order term for the $\omega$ meson, the
quantum effects generated due to the $\omega$ meson through the
present variational ansatz turn out to be zero.

We then write down the expressions for the thermodynamic quantities
including the quantum effects. The details regarding the formalism can
be found in earlier references \cite {mishra,hotnm,vecmass}.
The energy density, after carrying out the renormalisation procedures
for the baryonic and scalar meson sectors \cite{hotnm}, is obtained as
\begin{equation}
\epsilon_{\rm ren}=\epsilon_{\rm finite}^{(N)}
+\Delta \epsilon_{\rm ren}+\epsilon_\omega+\Delta \epsilon_\sigma,
\end{equation}
with
\begin{mathletters}
\begin{equation}
\epsilon_{\rm finite}^{(N)}=
\gamma (2\pi)^{-3}\int d \bfm k (k^2+{M^*}^2)^{1/2} 
(f_B +{\bar f}_B), 
\end{equation}
\begin{eqnarray}
\Delta \epsilon_{\rm ren} &=& -\frac{\gamma}{16\pi^2}
\biggl( {M^*}^4 \ln \Big (\frac{M^*}{M}\Big )
+M^3 (M-M^*)-\frac{7}{2} M^2 (M-M^*)^2 \nonumber\\
&& +  \frac{13}{3} M (M-M^*)^3 -
 \frac{25}{12} (M-M^*)^4 \biggr),
\end{eqnarray}
\begin{equation}
\epsilon_{\omega}=
g_\omega \omega_0 \rho_B^{\rm ren}-\frac {1}{2} m_\omega^2 \omega_0^2,
\end{equation}
\begin{eqnarray}
\Delta \epsilon_\sigma 
&=& \frac{1}{2} m_R^2 \sigma_0^2+ 3\lambda_R \sigma_0^4 
+\frac {M_\sigma^4}{64\pi^2}
\Biggl(\ln\Big(\frac{M_\sigma^2}{m_R^2}\Big)-\frac{1}{2} \Biggr)
-3\lambda_R I_f^2\nonumber\\
&& -\frac {M^4_{\sigma,0}}{64\pi^2}
\Biggl(\ln\Big(\frac{M_{\sigma,0}^2}{m_R^2}\Big)-\frac{1}{2} \Biggr)
+3\lambda_R I_{f0}^2,
\label{vph0}
\end{eqnarray}
\end{mathletters}

\noindent as the mean field result, contribution from the Dirac sea,
and contributions from the $\omega$ and $\sigma$ mesons, respectively.
We might note here that the quantum effects arising here from the
scalar meson sector through $\sigma$ meson condensates amounts to a
sum over a class of multiloop diagrams and, does not correspond to the
one meson loop approximation for scalar meson quantum effects
considered earlier \cite{serot,pal}.  In the above, $f_B$ and $\bar f
_B$ are the thermal distribution functions for the baryons and
antibaryons, given in terms of the effective energy,
$\epsilon^*(k)=(k^2+{M^*}^2)^{1/2}$,
and  the effective chemical potential, $\mu^{*}=\mu -g_\omega \omega_0$.
$M^{*}=M+g_\sigma \sigma_0$ is the effective nucleon mass and 
$\rho_B^{\rm ren}$ is the baryon number density after subtracting 
out the vacuum contribution.
The spin-isospin degeneracy factor is $\gamma=4$ for symmetric nuclear
matter. In the expression for the energy density arising from the scalar
meson sector, the field dependent effective sigma mass,
$M_\sigma(\beta)$, satisfies the gap equation in terms of the
renormalised parameters as
\begin{equation}
M_\sigma(\beta)^2=m_R^2+12\lambda_R\sigma_0^2+12\lambda_R I_f(M_\sigma(\beta)),
\label{mm2}
\end{equation}
where
\begin{equation}
I_f(M_\sigma(\beta))=\frac{M_\sigma(\beta)^2}{16\pi^2}
\ln \Big(\frac{M_\sigma(\beta)^2}{m_R^2} \Big)+
\frac{1}{(2\pi)^3}\int \frac {d\bfm k}{(\bfm k^2+ M_\sigma(\beta)^2)^{1/2}}
\frac {1}{e^{\beta\omega'(\bfm k,\beta)} -1}
\label{if}
\end{equation}
where $\omega'(\bfm k,\beta)=(\bfm k^2+M_\sigma(\beta)^2)^{1/2}$.  The
effective meson mass $M_\sigma$ is determined through a self
consistent solution of Eq. (\ref{mm2}).  In Eq. (\ref{vph0}),
$M_{\sigma,0}$ and $I_{f0}$ are the expressions as given by
Eqs. (\ref{mm2}) and (\ref{if}) with $\sigma_0=0$.  We might note here
that the gap equation given by Eq. (\ref{mm2}) is identical to that
obtained through resumming the daisy and superdaisy graphs
\cite{amhm,pi} and hence includes higher--order corrections from the
scalar meson field in a self-consistent manner.

Extremisation of the thermodynamic potential,
$\Omega =-p=\epsilon_{\rm ren} -\frac {\cal S}{\beta}-\mu {\rho_B}^{\rm ren}$,
with respect to the meson fields $\sigma_0$
and $\omega_0$ yields the self--consistency conditions for $\sigma_0$
(and hence for the effective nucleon mass $M^*=M+g_\sigma \sigma_0$)
and for the vector meson field $\omega_0$.

\section {In-medium properties for $\omega$ and $\rho$ vector mesons}

\subsection {Vacuum polarisation}

We now examine the medium modification to the masses and decay widths
of the $\omega$ and $\rho$ mesons including the quantum correction effects 
in hot nuclear matter in the relativistic random phase approximation. 
The interaction vertices for these mesons with nucleons are given as

\begin{equation}
{\cal L}_{\rm int}=g_{V}\Big (\bar \psi \gamma_\mu \tau^a \psi V_a^{\mu}
-\frac {\kappa_V}{2 M_N} \bar \psi \sigma_{\mu \nu} \tau^a \psi
\partial ^\nu V_a^\mu  \Big)
\label{lint}
\end{equation}
\noindent where $V_a^\mu=\omega^\mu$ or $\rho _a^\mu $, $M_N$ is the free
nucleon mass, $\psi$ is the nucleon field and $\tau_a=1$ or $\vec \tau $,
$\vec \tau$ being the Pauli matrices. $g_V$ and $\kappa_V$ correspond
to the couplings due to the vector and tensor interactions for the
corresponding vector mesons to the nucleon fields.  
The vector meson self energy is expressed 
in terms of the nucleon propagator, $G(k)$ modified by the quantum effects.
This is given as
\begin{equation}
\Pi ^{\mu \nu} (k)=-\gamma_I g_V^2 \frac {i}{(2\pi)^4}\int d^4 p\, 
{\rm Tr} \Big [ \Gamma_V^\mu (k) G(p) \Gamma_V^\nu (-k) G(p+k)\Big],
\end{equation}
where $\gamma_I=2$ is the isospin degeneracy factor for
nuclear matter, and $\Gamma_V^\mu (k)=\gamma^\mu \tau_a -
\frac {\kappa_V}{ 2 M_N}\sigma^{\mu \nu}$
represents the meson-nucleon 
vertex function.

The vector meson self energy can be written as the sum of two parts
\begin{equation}
\Pi^{\mu \nu}= \Pi^{\mu \nu}_F+ \Pi^{\mu \nu}_D.
\end{equation}
where $\Pi^{\mu \nu}_F$ is the contribution arising from the vacuum
polarisation effects, described by the coupling to the $N\bar N$
excitations and $\Pi^{\mu \nu}_D$ is the density dependent
contribution to the vector self energy. For the $\omega$
meson, the tensor coupling is generally small as compared to the
vector coupling to the nucleons \cite {hatsuda1}, and is
neglected in the present work. The Feynman part of the self energy, 
$\Pi ^ {\mu \nu}_F$ is divergent and needs renormalization. 
We use dimensional regularization to separate the divergent parts
and the renormalization condition 
$\Pi^{\rm ren}_F (k^2)|_{M_N^* \rightarrow M_N}=0$.
Because of the  tensor interaction, the vacuum self energy for
the $\rho$ meson is not renormalizable. We employ a phenomenological 
subtraction procedure \cite {hatsuda,hatsuda1} to extract the
finite part using the renormalization condition
$\left.\frac {\partial ^n \Pi _F^\rho (k^2)}{\partial (k^2)^n}\right|_{M_N^*
\rightarrow M_N}=0$, with ($n=0,1,2,\cdots \infty$). 

The effective mass of the vector meson is obtained by solving the equation,
with $\Pi =\frac {1}{3} \Pi^\mu _\mu$, 
\begin{equation}
k_0^2-m_V^2 + {\rm Re} \Pi (k_0,{\bfm k}=0) =0.
\label {omgrho}
\end{equation}

\subsection {Meson decay properties}

The decay width for the process $\rho \rightarrow \pi \pi$ is
calculated from the imaginary part of the self energy using the
Cutkosky rule, and in the rest frame of the $\rho$ meson 
is given by
\begin{equation}
\Gamma _\rho (k_0)=\frac {g_{\rho \pi \pi}^2}{48\pi} 
\frac {(k_0^2-4 m_\pi^2)^{3/2}}{k_0^2} \Bigg [
\Big (1+f(\frac {k_0}{2}) \Big )
\Big (1+f(\frac {k_0}{2}) \Big )
-f(\frac {k_0}{2}) f(\frac {k_0}{2}) \Bigg ]
\label {gmrho}
\end{equation}
where, $f(x)=[e^{\beta x}-1]^{-1}$ is the Bose-Einstein distribution 
function. The first and the second terms in the equation (\ref{gmrho})
represent the decay and the formation of the resonance, $\rho$.
The medium effects have been shown to play a very important role
for the $\rho$ meson decay width. In the calculation
for the $\rho$ decay width, the pion has been treated as free, 
and any modification of the pion propagator due to 
effects such as delta-nucleon hole excitation \cite {asakawa} 
to yield a finite decay width for the pion, have not
been taken into account. 
The coupling $g_{\rho \pi \pi}$ is fixed from the decay width
of $\rho$ meson  in vacuum ($\Gamma_\rho$=151 MeV) decaying 
to two pions.

For the nucleon-rho couplings, we use the vector and tensor couplings 
as obtained from the N-N forward dispersion relation
\cite{hatsuda1,sourav,grein}. With the couplings as described above, we
consider the temperature and density dependence
of the $\omega$ and $\rho$ meson properties in the hot nuclear matter
due to quantum correction effects. 

To calculate the decay width for the $\omega$ meson, we
write down the effective Lagrangian for the $\omega$ meson as
\cite {sakurai,gellmann,bali}
\begin{equation}
{\cal L}_\omega=-\frac {em_\omega^2}{g_\omega}\omega^\mu A_\mu
+\frac {g_{\omega \pi \rho}}{m_\pi}\epsilon _{\mu \nu \alpha  \beta}
\partial ^\mu \omega^\nu \partial ^\alpha {\rho}^{\beta}_i
{\pi_i}+
\frac {g_{\omega 3 \pi}}{m_\pi ^3}\epsilon _{\mu \nu \alpha  \beta}
\epsilon _{ijk}\omega ^\mu \partial ^\nu \pi ^i  \partial ^\alpha \pi ^j  
\partial ^\beta \pi ^k. 
\label{lomg}
\end {equation} 
In the above, the first term refers to the direct coupling of
the vector meson $\omega$ to the photon,  and hence to the dilepton
pairs, as given by the vector dominance model.
The decay width of the $\omega$ meson in vacuum is dominated
by the channel $\omega \rightarrow 3 \pi$.
In the medium, the decay width for $\omega \rightarrow 3\pi$
is given as
\begin{eqnarray}
\Gamma _{\omega \rightarrow 3 \pi} & = & \frac {(2\pi)^4}{2 k_0}
\int d^3 {\tilde p}_1   d^3 {\tilde p}_2   d^3 {\tilde p}_3\,
\delta ^{(4)} (P-p_1-p_2-p_3) |M_{fi}|^2
\nonumber \\ && \Big [(1+f(E_1))(1+f(E_2))(1+f(E_3))
-f(E_1)f(E_2)f(E_3) \Big ],
\end{eqnarray}
where $d^3 {\tilde p}_i=\frac {d^3 p_i}{(2\pi)^3 2 E_i}$, $p_i$ and
$E_i$'s are 4-momenta and energies for the pions, and $f(E_i)$'s
are their thermal distributions. The matrix element $M_{fi}$ has 
contributions from the channels 
$\omega \rightarrow \rho \pi \rightarrow 3\pi$ (described by the second
term in (\ref {lomg})) and the direct decay $\omega \rightarrow 3\pi$
resulting from the contact interaction (last term in (\ref {lomg}))
\cite{bali,weisezp,kaymak}.
For the $\omega \rho \pi$ coupling we take the value $g_{\omega \rho \pi}$
=2 according to Ref. \cite{sourav}, which is similar in value to the coupling
taken in \cite{weisezp}. We fix the point interaction coupling 
$g_ {\omega 3 \pi}$ 
by fitting the partial decay width $\omega \rightarrow 3 \pi$ in vacuum
(7.49 MeV) to be 0.24. The contribution arising from the direct decay 
is upto around 15 \% which is comparable to the results of 
\cite{bali,weisezp}. 

With the modifications of the vector meson masses in the hot and dense 
medium, a new channel becomes accessible, i.e., the decay mode
$\omega \rightarrow \rho \pi$ for $m^*_\omega > m^*_\rho +m_\pi$, 
which is not possible in the vacuum (since $m_\omega < m_\rho +m_\pi$).
The in-medium decay width for this process is found from the cutting
rules at finite temperature to be
\begin{equation}
\Gamma_{\omega \rightarrow \rho \pi}
=\frac {g_{\omega \pi \rho}}{32 \pi {m^*_\omega} ^3 m_\pi^2}
\lambda^{3/2}({m^*_\omega}^2,{m^*_\rho}^2,{m_\pi}^2)
\Big [\big(1+f(E_\pi)\big)\big(1+f(E^*_\rho)\big)-
f(E_\pi)f(E^*_\rho)\Big ]
\end{equation}
where $\lambda (x,y,z)=x^2+y^2+z^2-2(xy+yz+xz)$ arises from phase
space considerations, and $f$'s are the thermal distribution functions
for the pions and $\rho$ mesons. The first and second terms refer to
the emission and absorption of pion and $\rho$ meson. 

\section {Dilepton production }

The pion annihilation ($\pi^+\pi^- \rightarrow e^+ e ^-$) dominates
dilepton production in the low invariant mass region. The dropping of
the vector meson masses in the hot and dense matter also gives rise to
substantial contribution to the dilepton yield from the direct decays
of the vector mesons into dileptons ($\rho \rightarrow e^+ e ^-$,
$\omega \rightarrow e^+ e^-$).

The thermal production rate per unit four volume for lepton pairs
is related to the imaginary part of the one--particle irreducible
photon self energy by \cite{sourav,gale,lebellac}

\begin{equation}
\frac {d R}{d^4 q} = \frac {\alpha}{12 \pi^4} \cdot 
\frac {{\rm Im} \Pi ^{\mu} _{\mu} (q)}{q^2 (e^{\beta q_0} -1)},
\end{equation}
where $q^\mu=(q_0,{\bfm  q})$ is the 4-momentum of the virtual photon,
and $\alpha=e^2/4\pi$ is the fine structure constant.

\subsection {Pion Annihilation ( $\pi^+\pi^- \rightarrow e^+ e ^-$)}

The invariant mass distribution for the lepton pairs from pion
annihilation is given by
\begin{equation}
\frac{dR}{dM}=\frac {M^3}{2(2\pi)^4} \big ( 1-\frac {4 m_\pi^2}{M^2}
\big ) \int dM_T M_T dy\, \sigma (q_0, |{\bfm q}|) f(M_T \cosh y)
\end{equation}
where, $q_0=M_T \cosh y$, $|{\bfm q}|=\sqrt {q_0^2-M^2}$.
In the above, $\sigma (q_0, |{\bfm q}|)$ is the cross section
for the pion annihilation, and is given by

\begin{equation}
\sigma (q_0, |{\bfm q}|)=\frac {4 \pi \alpha ^2}{3M^2} \sqrt 
{1-4 m_\pi ^2/M^2} |F_\pi (q_0,|{\bfm q}|)|^2,
\end{equation}
where  $F_\pi(q)$ is the electromagnetic form factor of the pion.
The photon can also couple
to the hadrons through vector mesons, which is the basis of
the vector dominance model \cite{sakurai}. The photon can
thus interact with the pions through a vector meson, which must
be an isovector and is taken to be the $\rho$ meson.
The effective Lagrangian for the photon-pion-rho interaction
can then be written as  \cite{sourav}
\begin {equation}
{\cal L}_{\rm int} =
 - g_{\rho \pi \pi} \rho^\mu J_\mu -eA^\mu J_\mu -\frac {e}{2g_\rho}
F^{\mu \nu} \rho_{\mu \nu},
\label{lrho}
\end {equation}
\noindent where $\rho_{\mu \nu}$ is the field tensor for the $\rho$ field
and $eJ_\mu=ie (\pi^- \partial _\mu \pi^+ - \pi^+ \partial _\mu \pi^-)$. 
The coupling $g_{\rho \pi \pi}$ in the above can be determined from the
decay $\rho \rightarrow \pi \pi$, and the direct photon-rho
coupling $g_\rho$ is determined from the process $\rho \rightarrow  e^+ e^-$
\cite {connell}. These, including a finite decay width
for the $\rho$ meson, then  lead to the $\rho$-dominated pion form factor as
\begin {equation}
F_\pi (q^2)=1-q^2 \frac {g_{\rho \pi \pi}}{g_\rho} \frac {1}{q^2-m_\rho^2
+im_\rho \Gamma _\rho}.
\end {equation}
In the rest frame of the $\rho$ meson, the invariant mass 
distribution of the lepton pair of invariant mass $M$
from pion annihilation is given as 
\begin{equation}
\frac {dR}{dM}=\frac {\sigma (M)}{(2\pi)^4} M^4 T 
\Big (1-\frac {4 m_\pi^2}{M^2}\Big )
\sum _{n=1} ^ \infty \frac {K_1 (nM/T)}{n} 
\end{equation}
with the cross section $\sigma (M)$ for the pion annihilation,
given in terms of the pion form factor at $q^2=M^2$.
The summation over $n$ arises from the Bose-Einstein distribution function
for the $\rho$ meson. Since we are mostly in the regime $M/T \gg 1$, the first 
term in the series, corresponding to the Boltzmannian distribution function
is dominant, with quantum corrections being of the order of upto 10 \%.

\subsection {Vector meson decaying to dileptons}

A vector meson $V$ in a thermal bath can decay into leptons and
photons, described by the final state $|f \rangle$ say, which escape
the bath without thermalization.  With $q$ as the four momentum of the
nonhadronic state $|f \rangle$, a resonance peak should appear in the
invariant mass ($M$) plot for the number of lepton pair events, versus
$q^2=M^2$ at $M=m_V$, where $m_V$ is the mass of the resonance in the
thermal bath. Hence, the medium modifications of the vector mesons get
reflected in the dilepton spectra through the shift and broadening of
the peaks.

The invariant mass distribution of lepton pairs from the decays
of vector mesons is given by a generalization of the Breit Wigner
formula as \cite{weldon}
\begin{equation}
\frac {dN_f }{d^4 x d^4 q}=(2J+1) \frac {M^2}{4\pi^2}
\frac {\Gamma _{{\rm all} \rightarrow V}
\Gamma _{V \rightarrow f}^{\rm vac}(M)}{(M^2-m_V^2)^2 
+ (m_V \Gamma _{\rm tot})^2}.
\label{gbw}
\end{equation}
In the above, $dN_f$ is the number of lepton pair events, say
in the  space time and four momentum element $d^4 x d^4 q$,
$J$ being the spin of the resonance, $\Gamma _{\rm all \rightarrow V}$
is the formation width, $\Gamma _{\rm tot}$ is the total width
and $\Gamma _{V \rightarrow f}^{\rm vac}$ is the partial decay width
for the off-shell $V$ (i.e. of mass $M$) to go into the nonhadronic 
state $|f \rangle$. The total width 
$\Gamma _{\rm tot}=\Gamma _{V \rightarrow {\rm all}}-
\Gamma _{{\rm all} \rightarrow V}$
is the rate at which the particles equilibriate and relax
to chemical equilibrium.
The invariant mass distribution of the lepton pairs from  decays
of vector mesons propagating with a momentum ${\bfm q}$ is obtained 
from equation (\ref{gbw}) as 

\begin{equation}
\frac {dR}{dM} = \frac {2J+1}{2 \pi^2} M \int M_T d M_T dy
f(M_T \cosh y) \times  \Big [\frac {q_0 
\Gamma_{\rm tot}}{(M^2 -m_V^2 + {\rm Re} \Pi)^2+q_0^2 \Gamma_{\rm tot}^2}\Big]
q_0 \Gamma _{V \rightarrow e^+ e^-}^{\rm vac}
\end{equation}
In the rest frame of the vector meson, this reduces to 
\begin{equation}
\frac {dR}{dM} = \frac {2J+1}{\pi^2} M^2 T \sum _n \frac {K_1(nM/T)}{n}
\times \frac {m^*_V \Gamma_{\rm tot}/\pi}{(M^2 -{m^*_V} ^2)^2
+{m^*_V}^2 \Gamma_{\rm tot}^2}
m^*_V \Gamma _{V \rightarrow e^+ e^-}^{\rm vac}(M).
\end{equation}
In the above, $ \Gamma _{V \rightarrow e^+ e^-}^{\rm vac}(M)$ is the 
partial width for the leptonic decay for the off-shell vector 
meson $V$ with mass $M$, and is given as
\begin{equation}
\Gamma _{V \rightarrow e^+ e^-}^{\rm vac}(M)
=\frac {4 \pi \alpha ^2 M}{3 g_V^2}
\end{equation}
The coupling $g_V$ refers to the direct coupling of the vector meson
$V$ to the photon as given in the equations (\ref {lomg}) and (\ref
{lrho}) for the $\omega$ and $\rho$ mesons respectively. The
coupling of $\omega$ to the photon taken to be one third of the coupling 
of $\rho$ to the photon gives reasonable agreement with the
ratio of the observed partial widths $\Gamma (\rho \rightarrow e^+
e^-)/\Gamma (\omega \rightarrow e^+ e^-)$ \cite{connell}.  

\section {Results and Discussions}

In the following we present numerical results on the influence of
vacuum effects on hadronic masses, decay widths of vector mesons, and
dilepton emission from hot hadronic matter.  As described in the
previous sections, the study is based entirely on a hadronic scenario
and makes use of the original Walecka model, extended by a nonlinear
sigma self coupling.

In this work we concentrate on one particular aspect, namely on the
role of scalar vacuum effects. For each quantity studied, we compare
the results of the Relativistic Hartree Approximation (RHA), which was
already treated in Ref. \cite{sourav}, with those additional effects
arising from a scalar vacuum condensate.  The quartic self coupling
$\lambda_R$ is taken as a free parameter; we have used $\lambda_R =
1.8$ and $\lambda_R = 5$ as representative values.  The values of the
incompressibility corresponding to these self couplings are 401 MeV
(for $\lambda_R$=1.8) and 329 MeV (for $\lambda_R$=5). These values
are smaller than the mean field result of 545 MeV as well as the
relativistic Hartree result ($\lambda_R$=0) of 450 MeV. This is
because an increase of $\lambda_R$ leads to weaker potentials of the
$\sigma$ and $\omega$ mesons and to a softening of the equation of
state.

The parameters of the model are adjusted to fit nuclear matter, 
corresponding to a saturation density of $\rho_0$ =0.193/fm$^3$. 
The fitted values for the scaled coupling constants are
$C_S^2 = g_\sigma^2 M_N^2/m_\sigma^2 =$ 183.3, 167.5 and 137.9, and
$C_V^2 = g_\omega^2 M_N^2/m_\omega^2 =$ 114.7, 96.45, and 63.7 for
$\lambda_R =0$ (i.e., RHA), 1.8, and 5, respectively.

When discussing dilepton production, we do not attempt to give a
realistic description of specific experimental results. For this, the
dynamic space-time evolution for the hot hadronic matter in the course
of a nuclear collision would have to be addressed and additional
effects contributing to the dilepton spectrum would have to be
included. However, our study should help to elucidate, in which way
in-medium vacuum effects can have an influence on experimentally
observable quantities.

We first discuss the effect of quantum corrections on the in-medium
nucleonic properties for hot nuclear matter \cite{hotnm}. In fig. 1a,
we plot the temperature dependence of the nucleon mass for the baryon
density $\rho_B=0$ and the nuclear matter density $\rho = \rho_0$. The
quantum corrections arising from the sigma field have the effect of
softening the equation of state, and produce a higher value for the
effective nucleon mass as compared to the RHA. The effective nucleon
mass was seen to increase with the vacuum polarisation effects arising
from the baryons as compared to the MFT calculations
\cite{chin,mishra}. Though for $\rho_B=0$, the effective nucleon mass
decreases monotonically with temperature, for $\rho_B=\rho_0$, the
nucleon mass first increases and then decreases as a function of
temperature, which has also been observed earlier to be the trend for
higher densities \cite{li}.  The variation in the nucleon mass is very
slow up to a temperature $T\simeq$ 150 MeV, beyond which there is a
fast decrease.  Though qualitative features are similar for the
in-medium mass with inclusion of quantum effects from the scalar meson
sector, it is observed that such effects lead to a higher value for
the nucleon mass.

We next study the temperature and density dependence of the vector mesons 
in hot nuclear matter. The $\omega$ meson mass as a function of the
temperature for $\rho_B$=0 and $\rho_0$, is plotted in Fig. 1b.
In the Walecka model, the Dirac sea has been shown to
have a significant contribution over the Fermi sea, leading
to a substantial drop in the omega meson mass in nuclear matter
\cite{hatsuda,hatsuda1,sourav}. Specifically, at saturation density 
and at zero temperature, the decrease in the omega meson mass from the
vacuum value, is around 150 MeV in RHA, whereas with sigma quantum
effects, the drop in the mass reduces to around 117 MeV or 61 MeV for
the two values of $\lambda_R$ studied. A similar reduction in the
mass drop, as compared to the RHA was also observed when scalar field
quantum corrections were included within a one--loop approximation
\cite{serot,pal}, unlike the approximation adopted here, which is
self-consistent and includes multi-loop effects.  This indicates that
the quantum corrections do play an important role in the medium
modification of the vector meson masses. In the present work, it is
seen that the quantum corrections from the scalar mesons, over those
arising from the baryonic sector, have the effect of giving rise to a
higher value for the $\omega$ meson mass. This again is related to the
fact that the effective nucleon mass is higher when such quantum effects
are taken into account.

In figure \ref{figf2m}, we illustrate the medium modification for 
the $\rho$ meson mass with the vector and tensor couplings 
to the nucleons being fixed from the NN forward dispersion relation
\cite{hatsuda1,sourav,grein}. The values
for these couplings are given as $g_{\rho N}^2/4\pi$=0.55
and $\kappa_\rho$=6.1. We notice that the decrease in the
$\rho$ meson mass with increase in temperature is much sharper than
that of the $\omega$ meson. Such a behaviour of the $\rho$ meson 
undergoing a much larger medium modification was also observed
earlier \cite {sourav} within the relativistic Hartree approximation
for the nucleons. This indicates that the tensor coupling,
which is negligible for the $\omega$ meson, 
plays a significant role for the $\rho$ meson. 
With inclusion of quantum corrections from 
the scalar meson, the qualitative comparison between the  $\omega$
and $\rho$ vector mesons remains the same, though these effects 
are generally seen to lead to larger values for the vector meson masses.
We have also calculated the
contribution to the $\rho$ meson mass arising from 
pion loops through the $\rho \pi \pi$ coupling, but this was found
to lead to a marginal increase of at most 1\%, in agreement with
\cite{weise,gale,sourav1}. 

In figure \ref{figf3m}, the decay widths for the $\rho$ meson with and
without the scalar quantum corrections, over the relativistic Hartree
approximation, are considered. Such quantum effects lead to an
appreciable increase in the decay widths, which, as we shall see,
is observed as a broadening of the $\rho$-peak in the dilepton spectra.

In figure \ref{figf4m}, the decay width for the $\omega$ meson is seen
to have a positive contribution arising from the $\sigma$-meson
quantum effect as compared to RHA. The decay width is calculated 
as arising from the processes $\omega \rightarrow 3\pi$ as well as
$\omega \rightarrow \rho \pi$, a channel which becomes possible with
the mass modifications of the vector mesons in the medium. 
It is seen that the channel $\omega \rightarrow
\rho \pi$ opens for temperature greater than around
180 MeV for $\rho_B=0$, whereas it is observed even at zero temperature
at nuclear matter saturation density. We might note that 
the collisional modification to the $\omega$ width,
arising from the process $\omega \pi \rightarrow \pi \pi$
\cite{sourav,weise1}, 
has not been taken into account in the present calculations. 

The dilepton spectra for the low invariant mass regime, at T=200 MeV
are shown in figures \ref{figf5m} and \ref{figf6m} calculated for densities
$\rho_B=0$ and $\rho_B=\rho_0$. The total dilepton spectra along with
contributions arising from pion annihilation and direct decays of the
$\rho$ and $\omega$ mesons, are displayed in figures 5a and 6a.  The
effect of sigma condensates shifts the peak position for the vector
mesons to higher values as compared to RHA, and also results in a
broadening of the peaks in the dilepton spectra as can be seen from
figures 5b and 6b.  The $\rho$ and $\omega$ peaks are well separated
due to the fact that the $\rho$ meson mass gets much larger medium
modification as compared to the $\omega$ meson, which would not be 
the case in the absence of medium effects. 
The omega peak is seen to be smeared due to scalar
meson quantum corrections.  The broadening of the $\omega$ peak due to
$\sigma$-quantum effects is more clearly visible for $\rho_B=\rho_0$
than for zero density, in the dilepton spectra. The dominant
contributions to the dilepton enhancement are from pion annihilation
and $\rho$ meson decay, which are plotted in figures 5c and 6c in
comparison with the RHA results.  The effect of $\sigma$ condensates
leads to a higher value for the peak position as well as broadening of
the $\rho$-peak. The dilepton production rates due to the $\omega$
meson decay are displayed in figures 5d and 6d. It is observed that
the broadening of the $\omega$ peak due to scalar quantum corrections
is larger for $\rho_0$ than for zero density. This is a reflection of
the fact that there is a larger modification of the omega width due to
scalar quantum effects for higher densities.  There have been various
mechanisms under discussion which can lead to an increase in the
$\omega$ width.  It has been argued that the $\omega$ decay width can
increase by an order of magnitude due to the reduction in the pion
decay constant related with chiral symmetry restoration \cite
{pisarski}.  The medium modification due to scalar meson quantum
effects giving rise to higher values for the $\rho$ meson mass and
decay width are reflected in the dilepton spectra.

The medium modification for the vector meson masses depend more
sensitively on the density than on temperature. This can give rise 
to an enhancement of the dilepton production rate when matter is 
highly compressed at the initial stage of a heavy ion collision.
In the DLS experiments for A-A collisions at 1 AGeV,
there have been models that focus on density induced change
for the $\rho$ meson mass to describe the dilepton pair
production for invariant mass of dilepton pair, $M \ge 0.4$GeV
\cite {liko,wamb1}. Even though there is dilepton enhancement
arising from the density--modified $\rho$ mass, the results still
underestimate the DLS data by a factor of 2-3 \cite{wamb1}. 

The mass modifications for the vector meson masses with density are
plotted in figure \ref{figf7}, for temperature T= 100 MeV.  The trend
is a sharp decrease of the masses with density up to $\rho_B \simeq
\rho_0$, and then a rise for higher values of $\rho_B$.  The observed
trend is due to the competing effects of vacuum polarisation which
leads to a reduction in the mass, and the density dependent dressing
for the vector self energy causing an increase in the meson mass. Such
a behaviour of the vector meson mass for the $\omega$ meson was also
observed earlier \cite{jeans}.  A higher value of the scalar self
interaction coupling, $\lambda_R$, leads to a higher value for the
vector meson masses. The change in the $\rho$ meson mass for higher
densities is seen to be rather slow which means that there is a
delicate cancellation between the Feynman and the density dependent
parts of the vector self energy. As depicted in figure \ref{figf7}, the
$\omega$ meson mass seems to be more dominated by the density
dependent contributions for higher densities.  The decay widths are
plotted in figure \ref{figf8}. The slow change in $\rho$ meson mass
for higher densities is reflected in the in-medium $\rho$-decay width,
while the $\omega$ decay width $\Gamma_\omega$ grows strongly with
density and gets strongly enhanced by quantum effects from the $\sigma$ meson.

The dilepton spectra for $T=100$ MeV and $\rho_B=2\rho_0$ are plotted
in figures \ref{figf91} and \ref{figf92}.  These values should be
typical for the compression zone of a nuclear collision at energies of
a few GeV per nucleon. For example, for inclusive Ca+Ca collisions at 1 AGeV
studied by the DLS collaboration \cite {dls}, maximum temperatures and
densities in this range were extracted from a transport model \cite
{zhang}. In figure 9a, the total dilepton spectrum is plotted, for
$\lambda_R$=1.8, displaying the individual contributions. 
Figure 9b shows that the sigma quantum corrections lead
to a broadening of the peaks in the dilepton spectra, which is seen to
be more appreciable for the higher value of the scalar self coupling,
$\lambda_R$. The individual contributions to the dilepton spectra,
plotted in figure \ref{figf92}, illustrate the effect of $\sigma$
quantum corrections.

Finally, the density dependence for the dilepton spectra is shown for
temperature $T=100$ MeV in figures \ref{figf101} and \ref{figf102} for
couplings $\lambda_R$ as 1.8 and 5, respectively, in comparison with
the RHA results.  The vector meson masses decreasing with density up
to around twice the nuclear matter density and then slowly increasing
or staying constant, are reflected in the dilepton spectra through the
peak positions.  One might notice that the broadening of the peaks due
to $\sigma$- quantum effects is larger for higher densities and is
more pronounced for the $\omega$ peak.  The curves for densities twice
and four times nuclear matter density in figure \ref{figf102} are
coincident up to a value of the invariant mass of about 0.6 GeV. This
is due to the fact that the rho meson properties whose effects are
relevant for that mass region, remain almost constant for
$\lambda_R$=5. Beyond that however, they are different around the
invariant mass region for the omega meson. Due to substantial increase
in the decay width of the $\omega$ meson with density the $\omega$
peak is almost smeared out even at nuclear matter density and the
omega loses its quasi particle nature. The corrections from the
scalar meson quantum effects are thus seen to be appreciable.

\section{Summary}

To summarize, we have considered in the present paper, the
modifications of the vector meson ($\rho$ and $\omega$) properties due
to quantum correction effects in hot and dense nuclear matter and
their effect on dilepton spectra.  The baryonic properties, as
modified due to such effects, subsequently determine the vector meson
masses and decay widths in the hot and dense hadronic matter.  It has
been recently emphasized that the Dirac sea contribution dominates
over the Fermi sea, thus implying that the vacuum polarisation effects
arising from the baryonic sector do play an important role in the
vector meson properties in a medium.  We consider the hadronic
properties as modified arising due to the quantum effects from the
baryon and scalar meson sectors and investigate their effect on the
dilepton production rate.  The effect of the scalar meson quantum
corrections leads to an increase in the vector meson masses and decay
widths.  The effect on the omega meson decay width is particularly
more pronounced at higher densities, than in the RHA. The value for
$T=100$ MeV is modified from 67 MeV (for RHA) to 125 MeV (for
$\lambda_R$=1.8) and 251 MeV (for $\lambda_R$=5) for $\rho_B/\rho_0$=2,
and for $\rho_B/\rho_0$=4 the RHA value of 101 MeV is modified to 175
MeV and 370 MeV, for $\lambda_R$=1.8 and 5, respectively .

In the present work, we have studied the effect of quantum corrections
arising from the scalar meson over those resulting from the baryon
sector (given through the summing over baryonic tadpole diagrams in
the relativistic Hartree approximation). The quantum corrections from
the $\sigma$-meson give rise to a higher value for the vector meson
masses as compared to keeping only quantum effects from the
baryons. There is also an appreciable increase in the decay widths in
the presence of these quantum corrections, and this is reflected in
the invariant mass plot for dilepton spectra through the broadening of
the peaks.  For the $\omega$ meson, an appreciable broadening arising
due to sigma meson quantum effects leads to the disappearance of the
peak from the spectra for high densities when $\omega$ ceases to be
exist as a quasi particle. The density dependence of in-medium
properties of the vector mesons and their subsequent effects on the
dilepton spectra have been studied, which will be relevant for highly
compressed matter at the early stage of a heavy ion collision.

\section{Acknowledgements}
One of the authors (AM) acknowledges financial support from 
Alexander von Humboldt foundation when the work was initiated.
AM would like to thank S. Scherer, H. Weber, D. Dietrich
for discussions, and would like to acknowledge the Institut f\"ur
Theoretische Physik for warm hospitality and DFG (STO 161/8-1)
for financial support.

\begin{figure}
\psfig{file=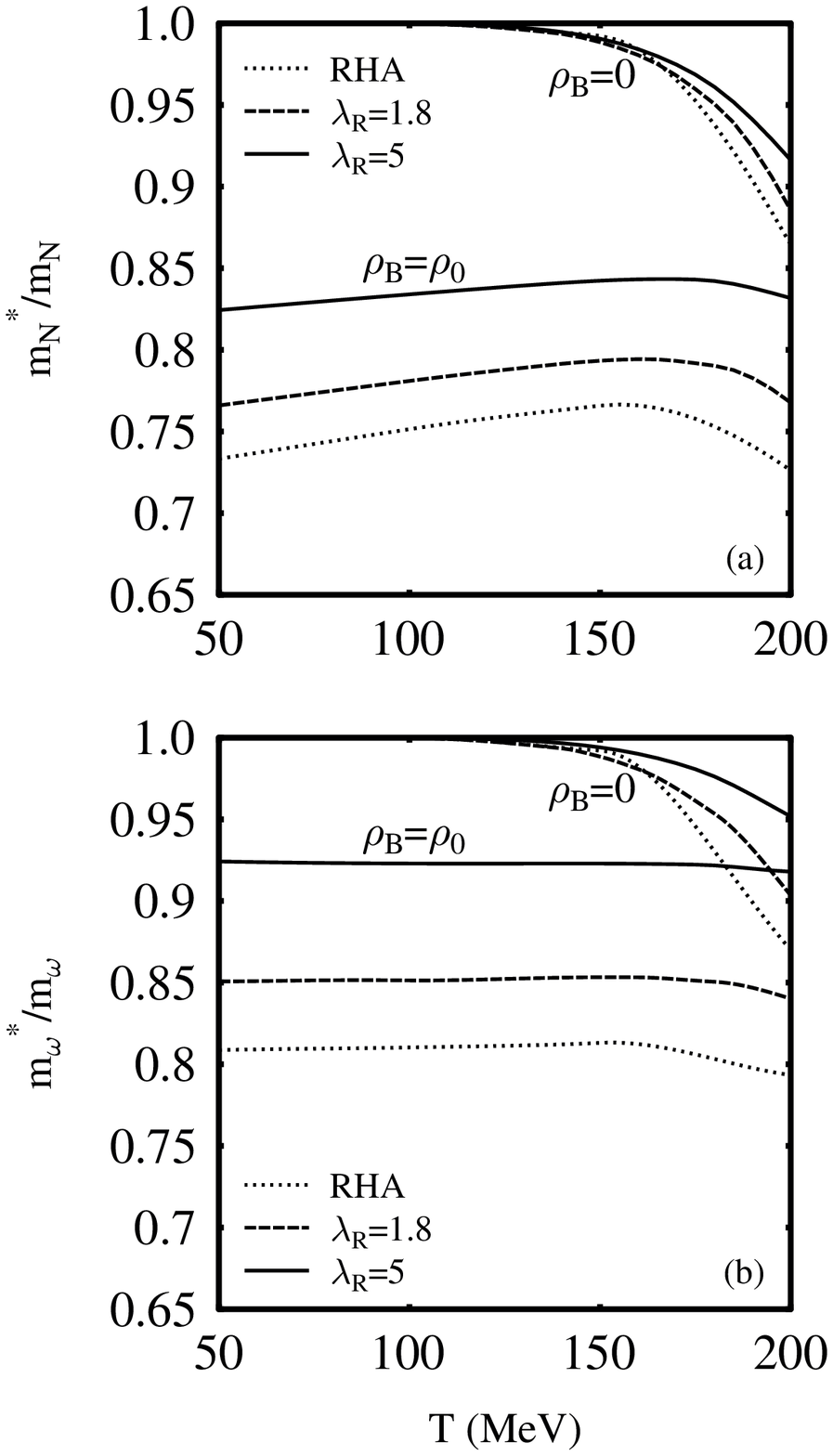,width=24cm,height=20cm}
\caption{Effective nucleon and omega meson masses as functions 
of the temperature
for zero and nuclear matter density for RHA, $\lambda_R=1.8$ and 
$\lambda_R=5$. The quantum corrections from the scalar meson sector
leads to an increase in the in-medium masses.}
\label{figf1m}
\end{figure}
\begin{figure}
\psfig{file=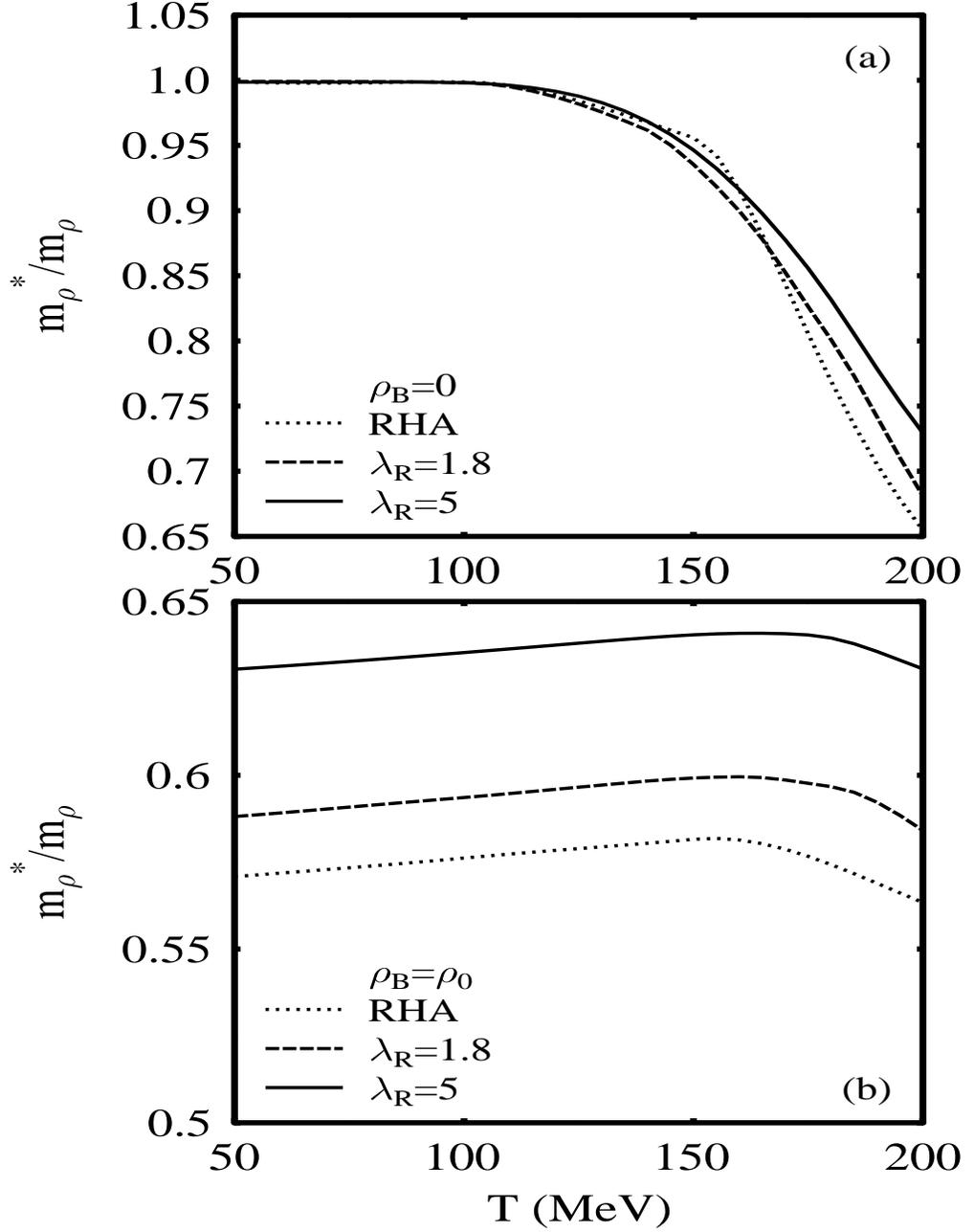,width=24cm,height=20cm}
\caption{Effective $\rho$ meson mass as a function of the temperature
for zero and nuclear matter densities for the three cases RHA, $\lambda_R=1.8$ 
and $\lambda_R=5$. 
The quantum corrections from the scalar meson sector 
increases the in-medium mass.}
\label{figf2m}
\end{figure}
\begin{figure}
\psfig{file=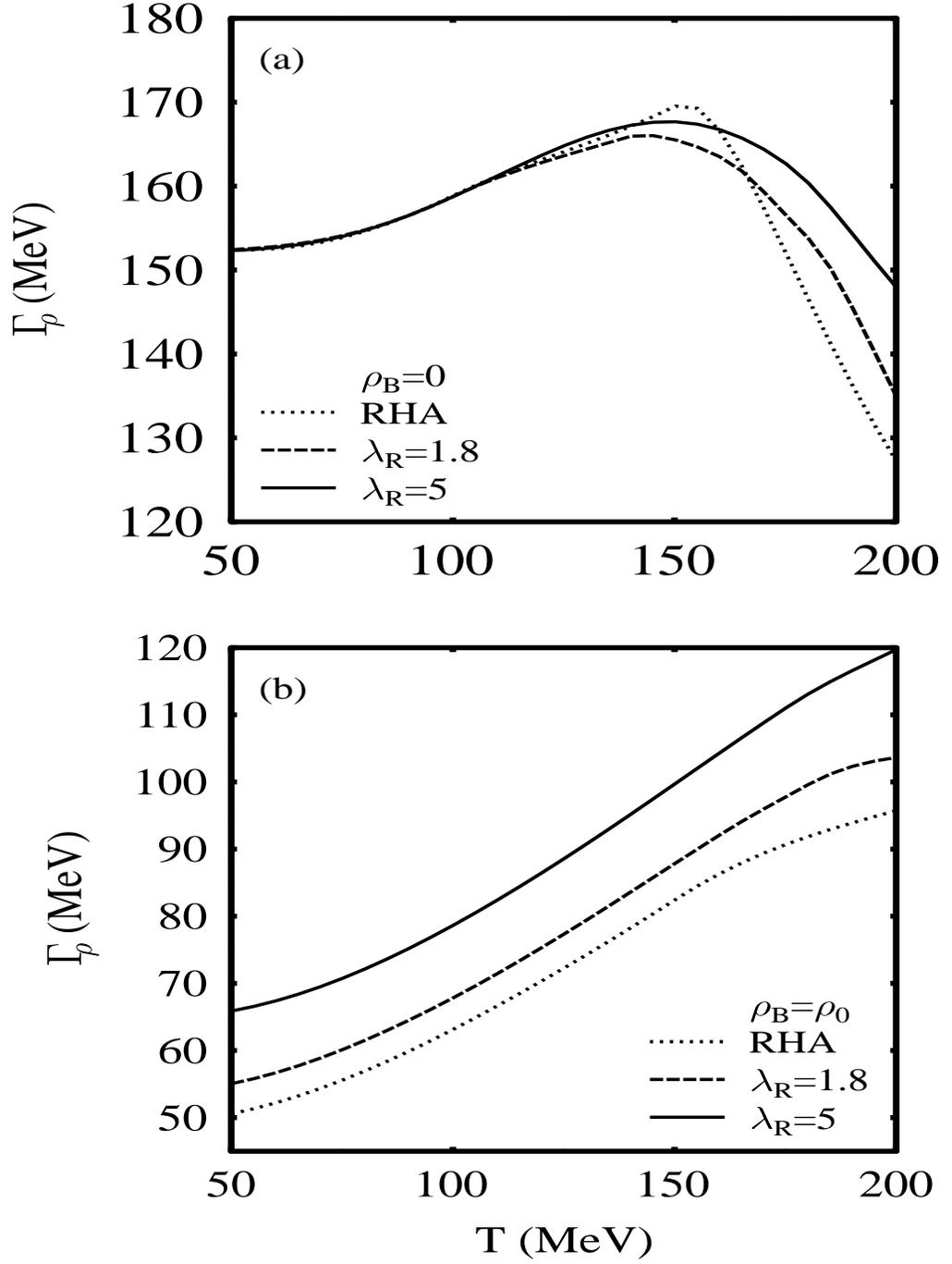,width=24cm,height=20cm}
\caption{In-medium decay width for the $\rho$ meson as a function 
of temperature and density.
} 
\label{figf3m}
\end{figure}
\begin{figure}
\psfig{file=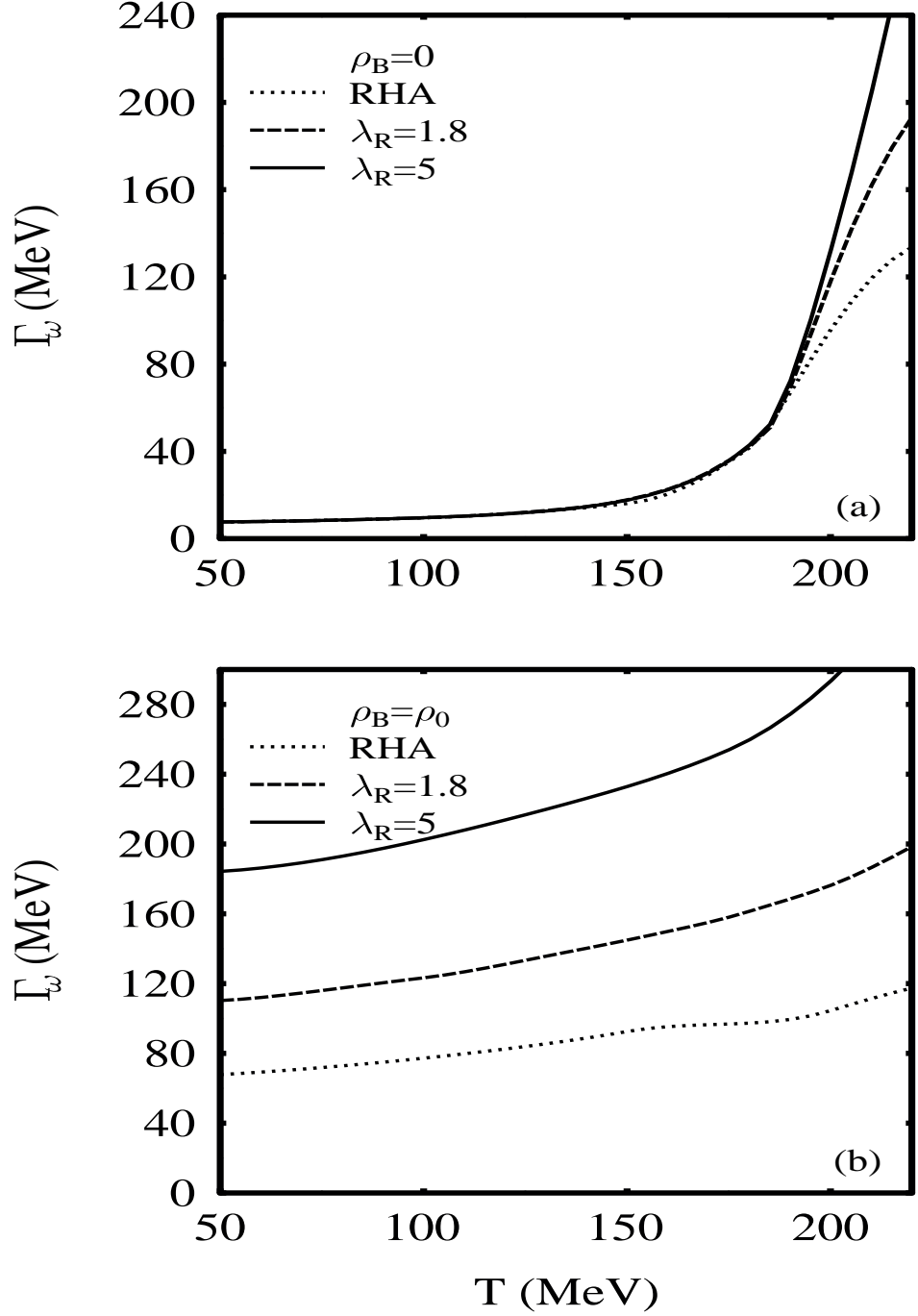,width=24cm,height=20cm}
\caption{In-medium decay width for the $\omega$ meson
arising from the channels $\omega \rightarrow 3\pi$
and $\omega \rightarrow \rho \pi$, the latter becoming accessible
because of the medium modification of the meson masses.}
\label{figf4m}
\end{figure}
\begin{figure}
\psfig{file=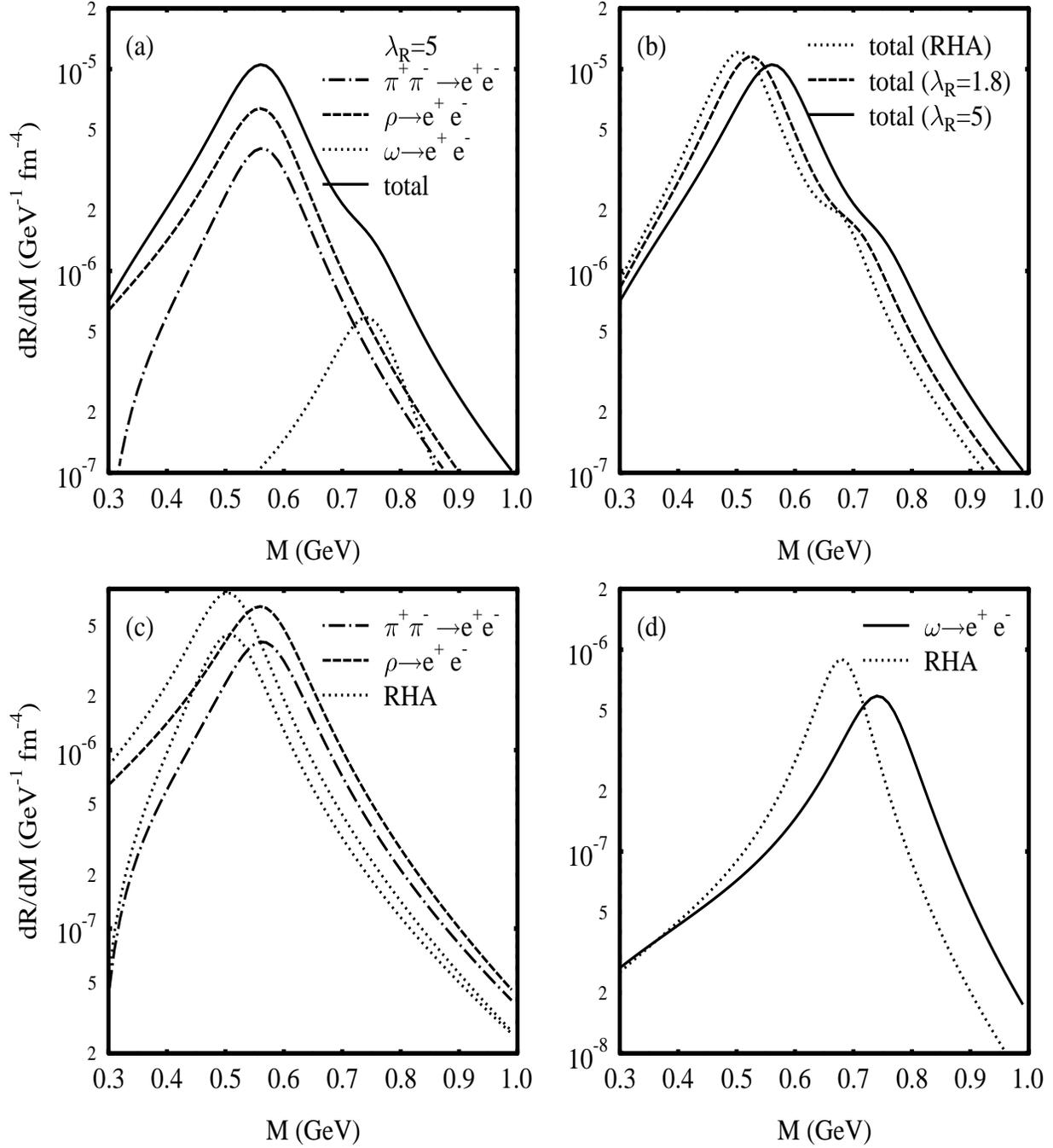,width=16cm,height=20cm}
\caption{Total dilepton production rate, as well as the individual
contributions from the pion annihilation and direct decays of the
$\rho$ and $\omega$ mesons are shown, with and without scalar meson 
quantum effects for $\rho_B=0$ and $T=200$ MeV. One observes
a positive shift as well as a broadening of the peaks due to 
$\sigma$ quantum effects.}
\label{figf5m}
\end{figure}
\begin{figure}
\psfig{file=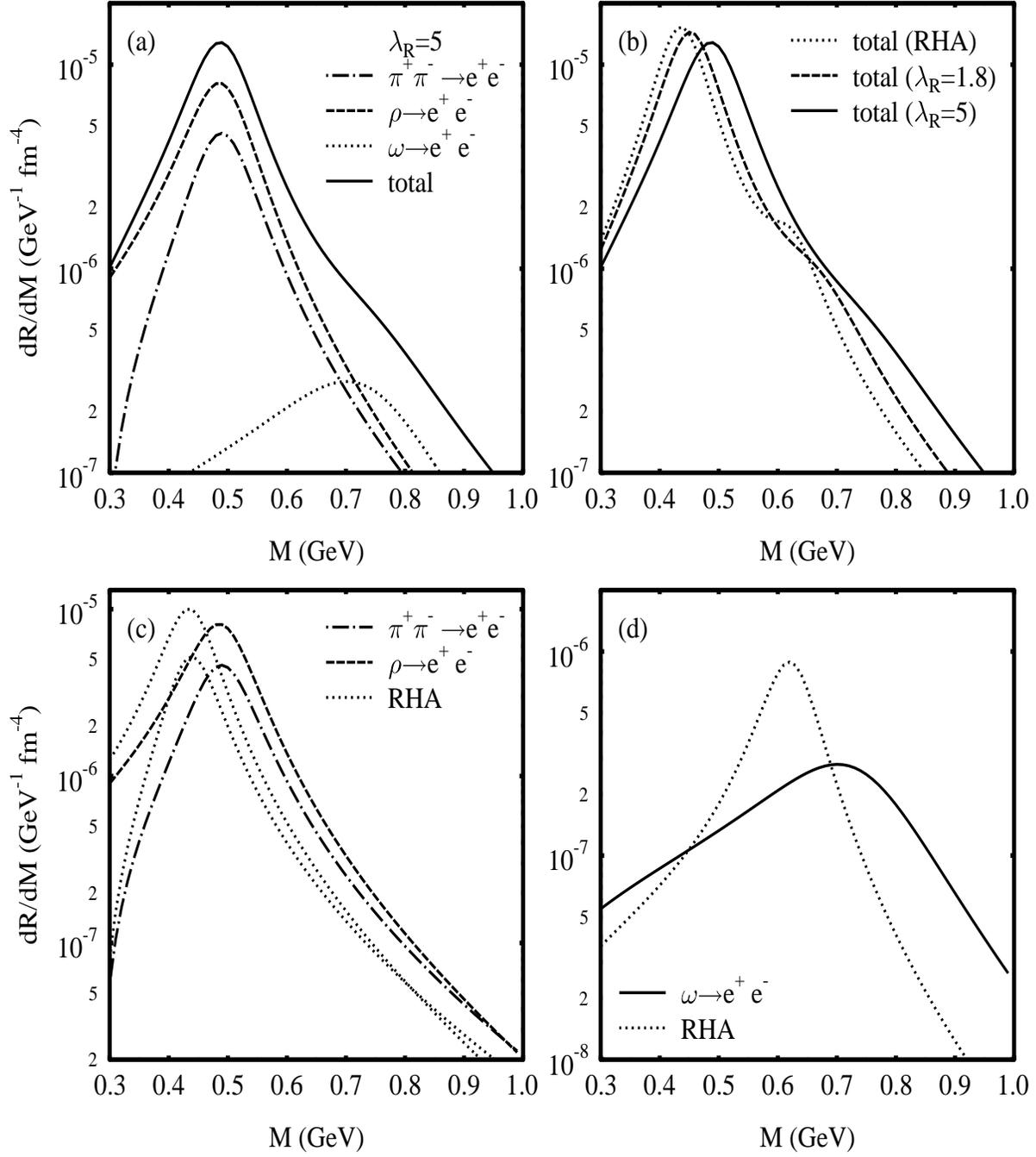,width=16cm,height=20cm}
\caption{Same as Fig. \ref{figf5m} for $\rho_B=\rho_0$} 
\label{figf6m}
\end{figure}
\begin{figure}
\psfig{file=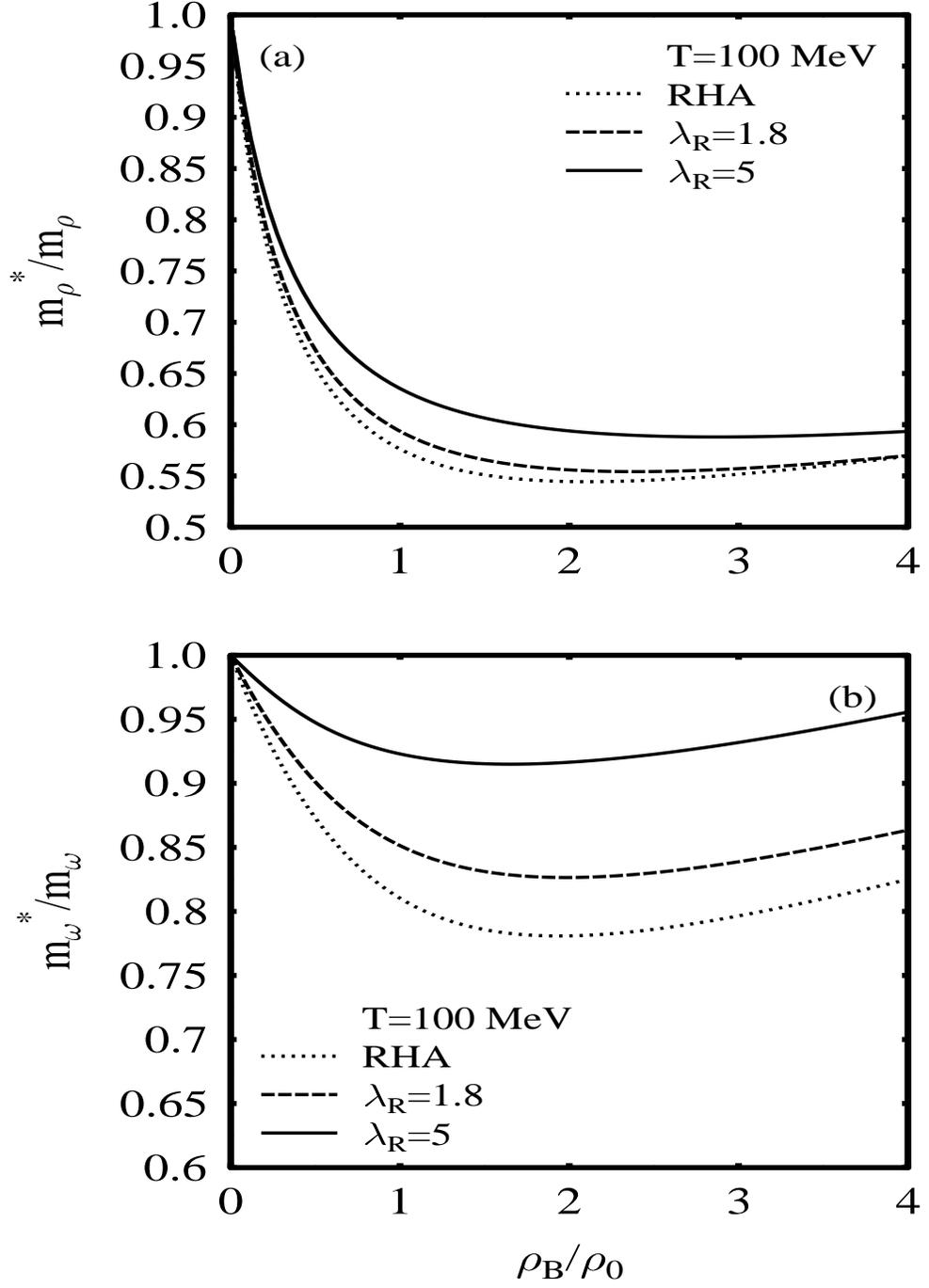,width=24cm,height=20cm}
\caption{Density dependence of the $\rho$ and $\omega$ meson masses
for $T=100$ MeV. The masses are seen to depend more sensitively on density than on 
temperature.
The $\sigma$ quantum effects lead to an increase in the in-medium masses.}
\label{figf7}
\end{figure}
\begin{figure}
\psfig{file=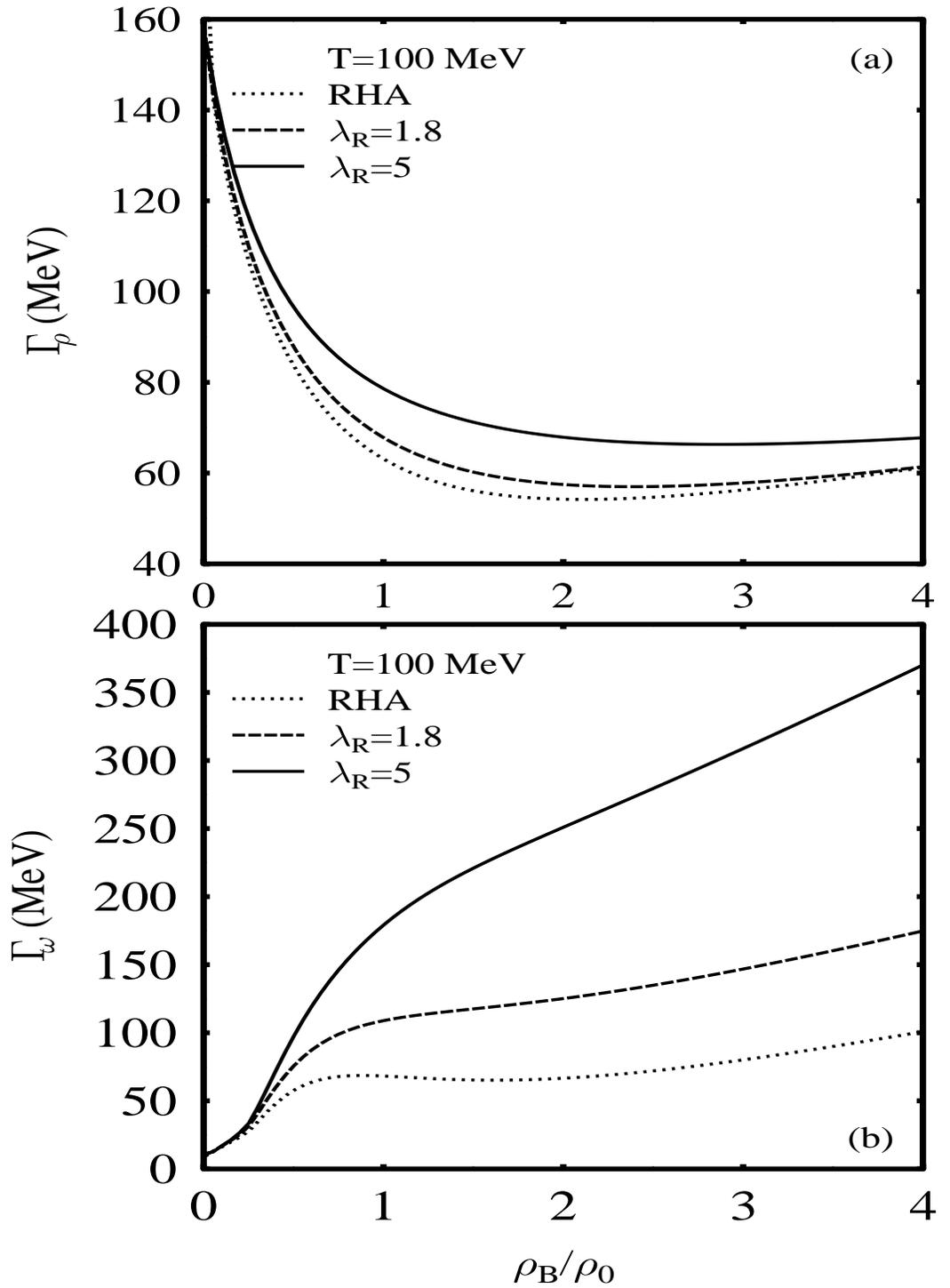,width=24cm,height=20cm}
\caption{Density dependence of the $\rho$ and $\omega$ meson widths.
With growing strength of the scalar self coupling, $\lambda_R$, the widths
increase appreciably.} 
\label{figf8}
\end{figure}
\begin{figure}
\psfig{file=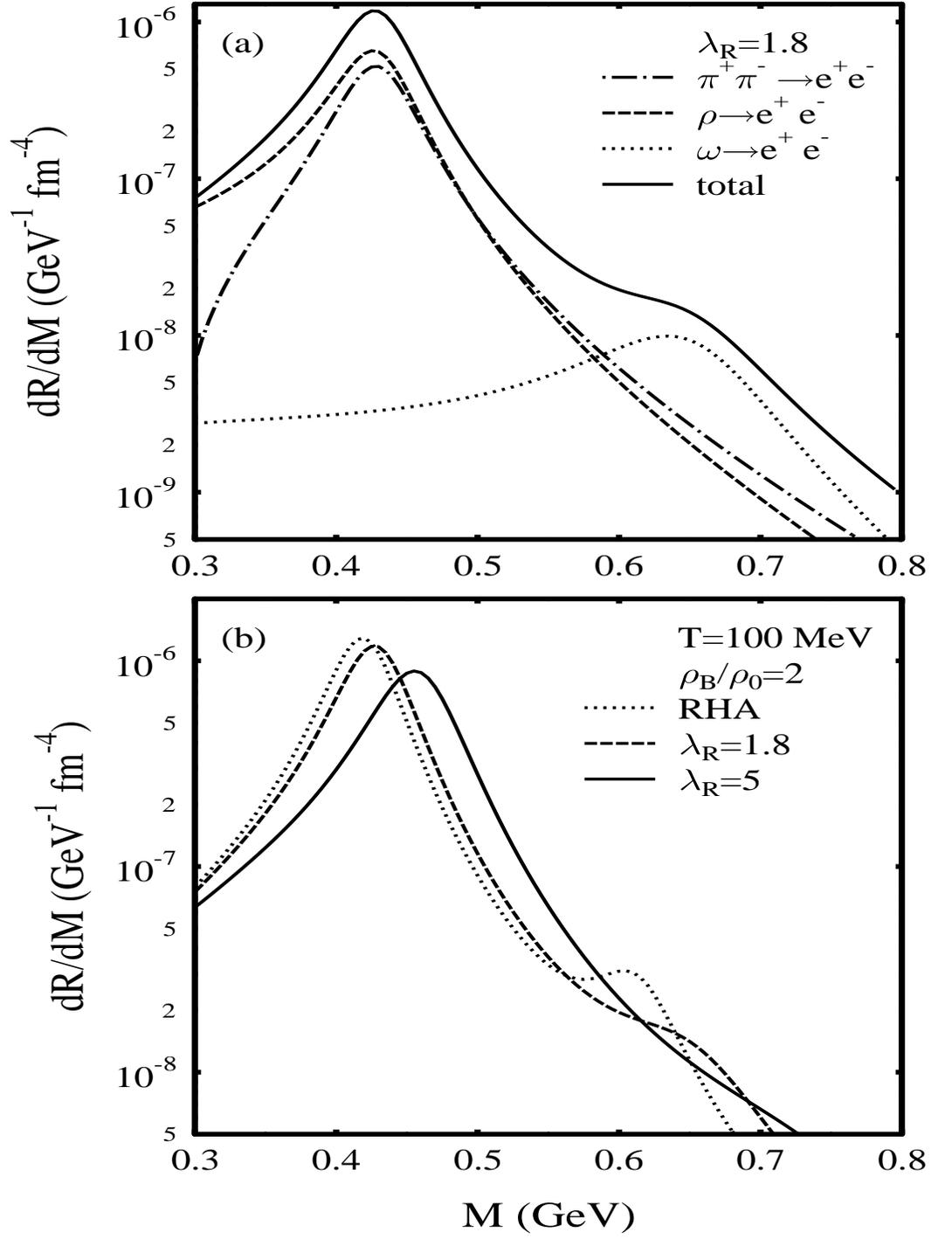,width=24cm,height=20cm}
\caption{Dilepton spectra for $T=100$ MeV and $\rho_B=2\rho_0$. The quantum
effects from $\sigma$ lead to a positive shift and broadening of the 
$\rho$ and $\omega$ peaks. For $\lambda_R = 5$ the $\omega$ peak is no longer
visible.}
\label{figf91}
\end{figure}
\begin{figure}
\psfig{file=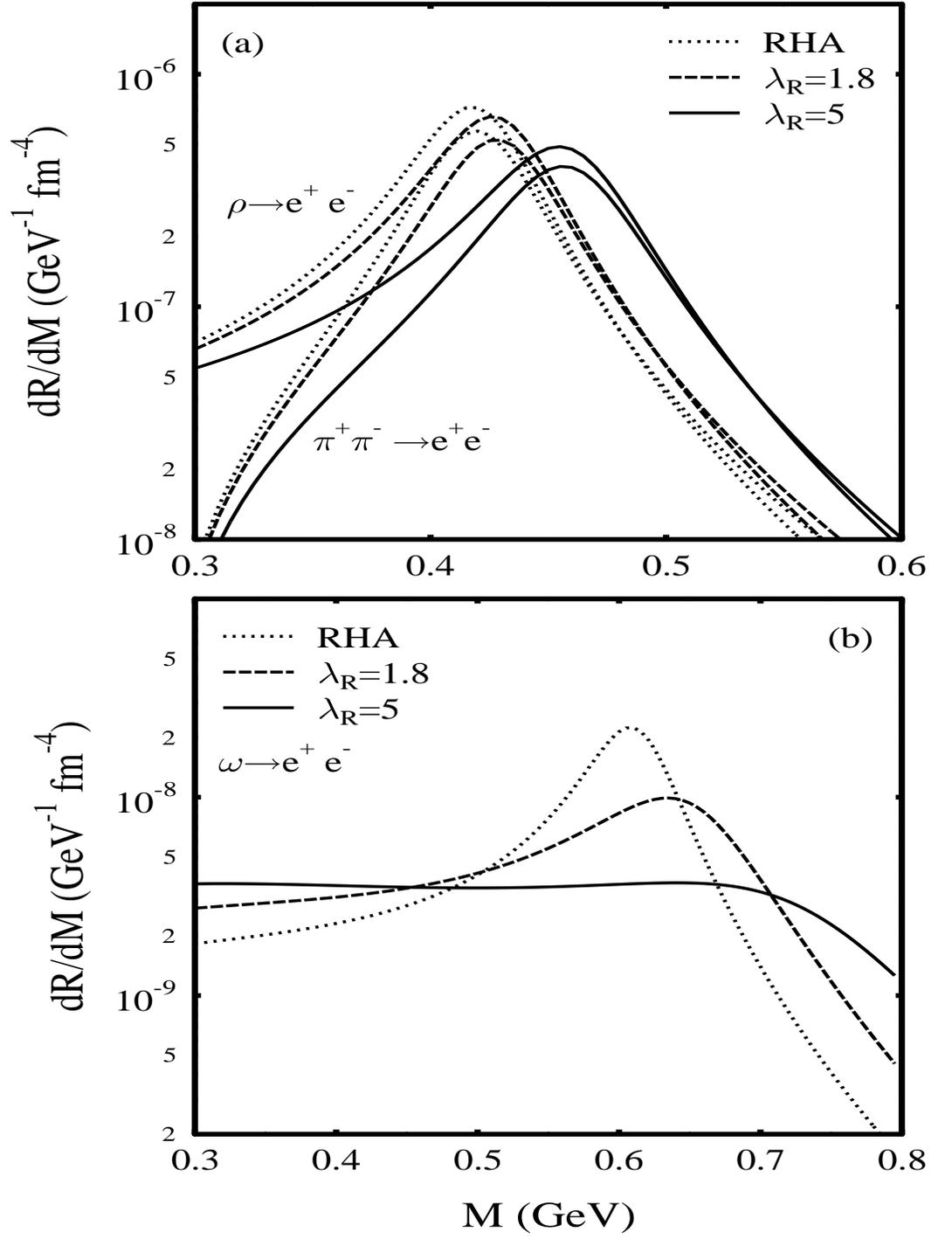,width=24cm,height=20cm}
\caption{Contributions from the processes of pion annihilation and
decay of vector mesons to the dilepton spectra. For the higher value
of $\lambda_R$ the $\omega$ meson peak is completely smeared out.
}
\label{figf92}
\end{figure}
\begin{figure}
\psfig{file=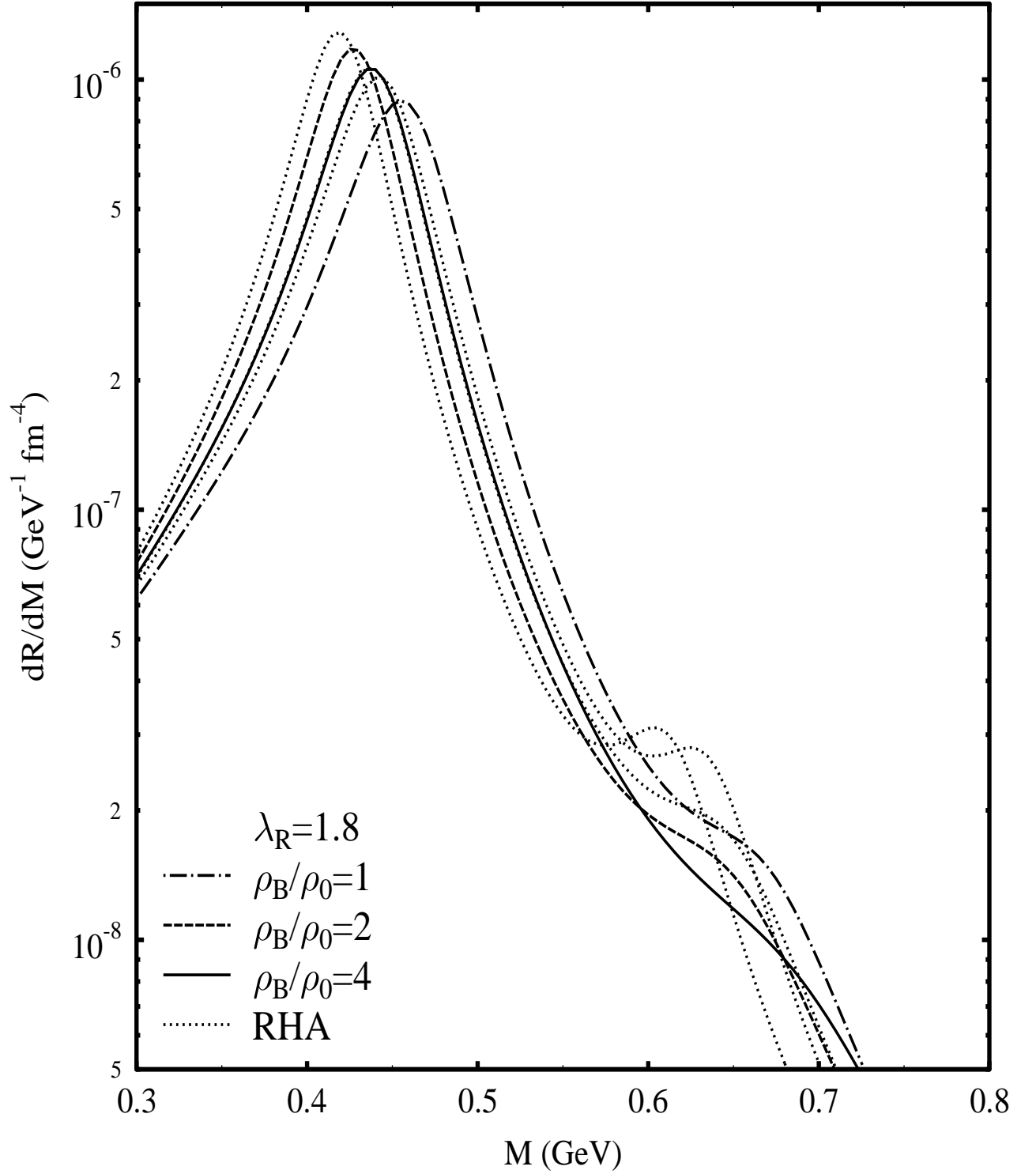,width=18cm,height=22cm}
\caption{Density dependence of dilepton spectra peak positions
and widths for $T=100$ MeV. Scalar quantum effects lead to a
broadening of the peaks. For the $\omega$ meson, the broadening is
more pronounced for higher densities.
}
\label{figf101}
\end{figure}
\begin{figure}
\psfig{file=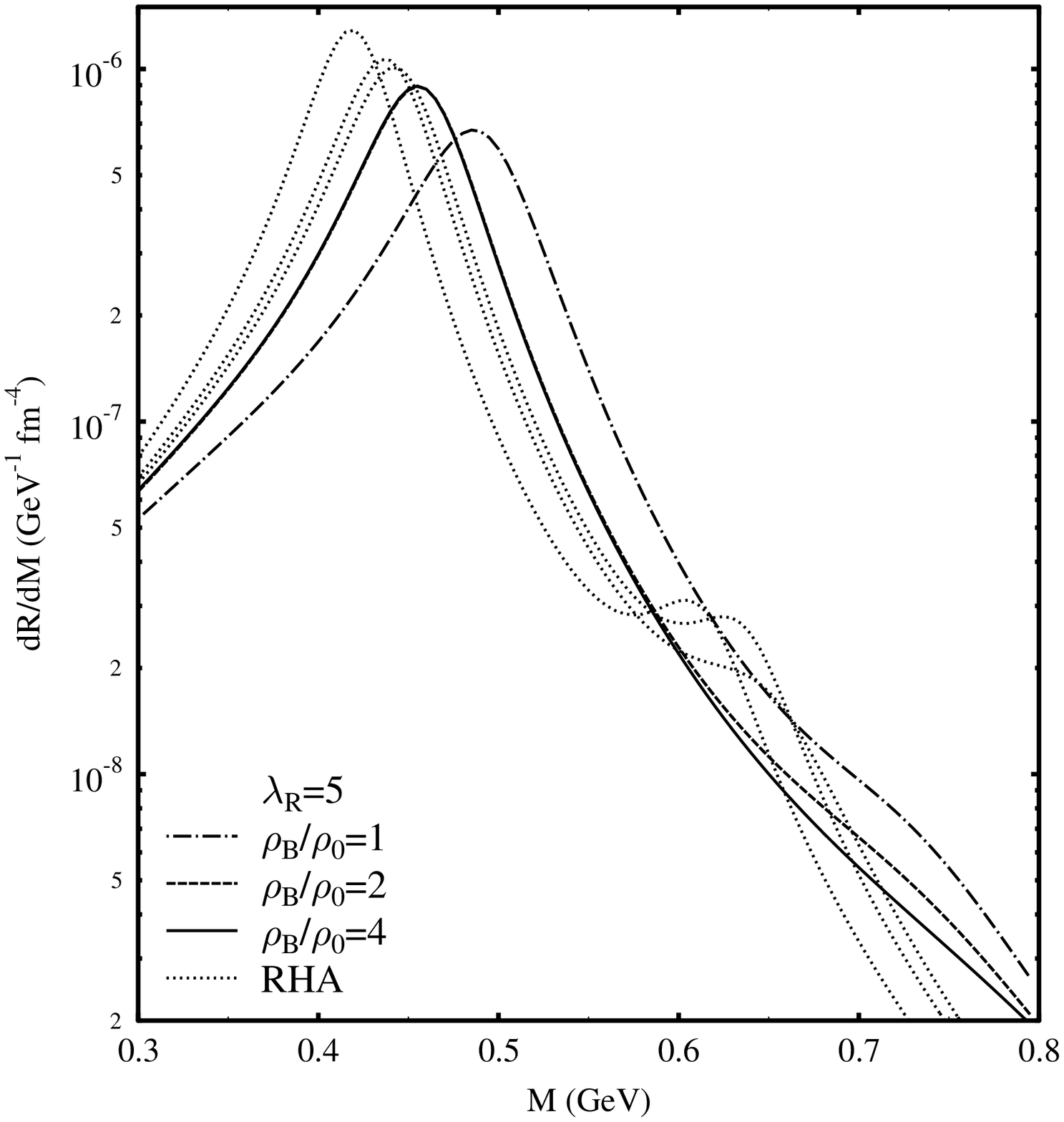,width=18cm,height=22cm}
\caption{Same as Fig. \ref{figf101} for $\lambda_R=5$. 
This leads to the disappearance of the $\omega$ peak even at nuclear
saturation density.}
\label{figf102}
\end{figure}

\begin{thebibliography}{99}
\bibitem{helios} N. Masera for the HELIOS-3 collaboration,
Nucl. Phys. {\bf A 590}, 93c (1995).
\bibitem {ceres}G. Agakichiev et al (CERES collaboration),
Phys. Rev. Lett. {\bf 75}, 1272 (1995); 
G. Agakichiev et al (CERES collaboration), Phys. Lett. {\bf B 422},
405 (1998); G. Agakichiev et al (CERES collaboration), Nucl. Phys. 
{\bf A 661}, 23c (1999).
\bibitem {dls} R. J. Porter et al (DLS collaboration), Phys. Rev.
Lett. {\bf 79}, 1229 (1997); W. K. Wilson et al (DLS collaboration),
Phys. Rev. {C 57}, 1865 (1998).
\bibitem {rhic} D. P. Morrison (PHENIX collaboration), Nucl. Phys.
{\bf A 638}, 565c (1998).
\bibitem {hades} J. Stroth (HADES collaboration), Advances 
Nuclear Dynamics {\bf 5}, 311 (1999). 
\bibitem {brown} G. E. Brown and M. Rho, Phys. Rev. Lett. {\bf 66}, 2720
(1991).
\bibitem{hat} T. Hatsuda and Su H. Lee, Phys. Rev. {\bf C 46},
R34 (1992); T. Hatsuda, S. H. Lee and H. Shiomi, Phys. Rev. {\bf C 52},
3364 (1995).
\bibitem{jin} X. Jin and D. B. Leinweber, Phys. Rev. {\bf C 52}, 3344 
(1995); T. D. Cohen, R. D. Furnstahl, D. K. Griegel and X. Jin,
Prog. Part. Nucl. Phys. {\bf 35}, 221 (1995); R. Hofmann, Th. Gutsche,
A. Faessler, Eur. Phys. J. {\bf C 17}, 651 (2000); S. Mallik and
K. Mukherjee, Phys. Rev. {\bf D 58}, 096011 (1998).
\bibitem {samir} S. Mallik and A. Nyffeler, Phys. Rev. {\bf C 63},
065204 (2001).
\bibitem{weise} F. Klingl, N. Kaiser, W. Weise, Nucl. Phys. {\bf A 624},
527 (1997).
\bibitem{ernst} C. Ernst, S. A. Bass, M. Belkacem, H. St\"ocker
and W. Greiner, Phys. Rev. {\bf C 58}, 447 (1998).
\bibitem{hatsuda} H. Shiomi and T. Hatsuda, Phys. Lett. {\bf B 334},
281 (1994).
\bibitem {hatsuda1} T. Hatsuda, H. Shiomi and H. Kuwabara, Prog. Theor.
Phys. {\bf 95}, 1009 (1996).
\bibitem{jeans} H.-C. Jeans, J. Piekarewicz and A. G. Williams,
Phys. Rev. {\bf C 49}, 1981 (1994); K. Saito, K. Tsushima, A. W. Thomas,
A. G. Williams, Phys. Lett. {\bf B 433}, 243 (1998).
\bibitem{sourav} Jan-e Alam, S. Sarkar, P. Roy, B. Dutta-Roy and 
B. Sinha, Phys. Rev. {\bf C 59}, 905 (1999).
\bibitem {wambach} R. Rapp and J. Wambach, Adv. Nucl. Phys.
{\bf 25}, 1 (2000).
\bibitem {koch} C. M. Ko, V. Koch and G. Q. Li, Ann. Rev. Nucl. Sci.
{\bf 47}, 505 (1997).
\bibitem {temp} Jan-e Alam, P. Roy, S. Sarkar and B. Sinha,
nucl-th/0106038;  Jan-e Alam, S. Sarkar, P. Roy, T. Hatsuda
and B. Sinha, Ann. Phys. {\bf 286}, 159 (2000); F. Karsch, 
E. Laermann, P. Petreczky, S. Stickan,
I. Wetzorke, hep-lat/0110208.
\bibitem {dens} K. Redlich, J. Cleymans, V. V. Goloviznin, in
Proceedings on NATO Advanced Workshop on Hot Hadronic Matter:
Theory and Experiment, N.Y. Plenum Press, 1995, 562p;
A. Dumitru, D. H. Rischke, Th. Sch\"onfeld, L. Winckelmann,
H. St\"ocker and W. Greiner, Phys. Rev. Lett. {\bf 70},
2860 (1993); P. Jaikumar, R. Rapp and I. Zahed, hep-ph/0112308;
Song Gao, Ru-Keng Su, Xue- Qian Li, Comm. Theor. Phys. {\bf 28},
207 (1997); D. Dutta, K. Kumar, A. K. Mohanty, R. K. Choudhury,
Phys. Rev. {\bf C 60}, 014905, 1999.
\bibitem {serot} B. D. Serot and J. D. Walecka, Adv. Nucl. Phys. {\bf 16}, 
1 (1986).
\bibitem {chin} S. A. Chin, Ann. Phys. {\bf 108}, 301 (1977); 
M. Asakawa, C. M. Ko, P. Levai and X. J. Qiu, Phys. Rev. {\bf C 46}, R1159 
(1992).
\bibitem {pal} S. Pal, Song Gao, H. St\"ocker and W. Greiner,
Phys. Lett. {\bf B 465}, 282 (1999).
\bibitem {mishra} A. Mishra, P. K. Panda, S. Schramm, J. Reinhardt
and W. Greiner, Phys. Rev. {\bf C 56}, 1380 (1997).
\bibitem{amhm} A. Mishra and H. Mishra, J. Phys. {\bf G 23}, 143 (1997).
\bibitem{pi} S.Y. Pi and M. Samiullah, Phys. Rev. {\bf D 36}, 3121 (1987);
G. A. Camelia and S.Y. Pi, Phys. Rev. {\bf D 47}, 2356 (1993).
\bibitem{hotnm} A. Mishra, P. K. Panda and W. Greiner, 
Jour. Phys. {\bf G 27}, 1561 (2001).
\bibitem{shm} A. Mishra, P. K. Panda and W. Greiner, 
Jour. Phys. {\bf G 28}, 67 (2002).
\bibitem{vecmass} A. Mishra, J. C. Parikh and W. Greiner,
Jour. Phys. {\bf G 28}, 151 (2002).
\bibitem{gale} C. Gale and J. I. Kapusta, Nucl. Phys. {\bf B 357},
 65 (1991).
\bibitem {weise1} R. A. Schneider, T. Renk and W. Weise, Nucl. Phys.
{\bf A 698}, 428 (2002); T. Renk, R. A. Schneider and W. Weise, 
Phys. Rev. {\bf C 66}, 014902 (2002). 
\bibitem{tfd} H. Umezawa, H. Matsumoto and M. Tachiki,
{\em Thermofield Dynamics and Condensed States} 
(North-Holland, Amsterdam, 1982).
\bibitem{asakawa} M. Asakawa, C. M. Ko, P. Levai and X. J. Qiu,
Phys. Rev. {\bf C 46}, R1159 (1992); M. Herrmann, B. L. Friman
and W. N\"orenberg, Nucl. Phys. {\bf A 560}, 411 (1993);
G. Chanfray and P. Shuck, Nucl. Phys. {\bf A 545}, 271c (1992). 
\bibitem{grein} W. Grein, Nucl. Phys. {\bf B 131}, 255 (1977);
W. Grein and P. Kroll, Nucl. Phys. {\bf A 338}, 332 (1980).
\bibitem{sakurai} J. J. Sakurai, {\em Currents and Mesons} (The University
od Chicago Press, Chicago, 1969).
\bibitem{gellmann} M. Gell-Mann, D. Sharp, and W. D. Wagner, Phys.
Rev. Lett. {\bf 8}, 261 (1962). 
\bibitem{bali} B. A. Li, Phys. Rev. {\bf D 52}, 5165 (1995).
\bibitem{weisezp} F. Klingl, N. Kaiser and W. Weise, Z. Phys. {\bf A 356},
193 (1996).
\bibitem{kaymak} \"O. Kaymakcalan, S. Rajeev and I. Schechter, Phys.
Rev. {\bf D 30}, 594 (1984).
\bibitem{lebellac} M. Le Bellac, {\em Thermal Field Theory} (Cambridge 
University Press, New York, 1996).
\bibitem{connell} H. B. Connell, B. C. Pearce, A. W. Thomas and 
A. G. Williams, Prog. Part. Nucl. Phys. {\bf 39}, 201 (1997).
\bibitem{weldon} H. A. Weldon, Ann. Phys. (N. Y. ) {\bf 228}, 43
(1993).
\bibitem {li} G. Q. Li, C. M. Ko, G. E. Brown, Nucl. Phys. 
{\bf A 606}, 568 (1996).
\bibitem{sourav1} S. Sarkar, Jan-e Alam, P. Roy, A.K. Dutt-Mazumder,
B. Dutta-Roy and B. Sinha, Nucl. Phys. {\bf A 634}, 206 (1998). 
\bibitem {pisarski} R. D. Pisarski, hep-ph/9503330.
\bibitem{liko} G. Chanfrey and P. Schuck, Nucl. Phys. {\bf A 555},
329 (1993); G. Q. Li and C. M. Ko, Nucl. Phys. {\bf A 582}, 731
(1995). 
\bibitem{wamb1} E. L. Bratkovskaya, W. Cassing, R. Rapp and
J. Wambach, Nucl. Phys. {\bf A 634}, 168 (1998).
\bibitem{zhang} J. Zhang and C. Gale, Phys. Rev. {\bf C 50}, 1617
(1994).
\end{thebibliography}
\end{document}